# Market Implications of Alternative Operating Reserve Modeling in Wholesale Electricity Markets

Hamid Davoudi, *Student Member, IEEE*, Fengyu Wang, *Senior Member, IEEE*, Yonghong Chen, *Fellow, IEEE*, Di Shi, *Senior Member, IEEE*, Alinson Xavier, Feng Qiu

*Abstract--* Pricing and settlement mechanisms are crucial for efficient resource allocation, investment incentives, market competition, and regulatory oversight. In the United States, Regional Transmission Operators (RTOs) adopts a uniform pricing scheme that hinges on the marginal costs of supplying additional electricity. This study investigates the pricing and settlement impacts of alternative reserve constraint modeling, highlighting how even slight variations in the modeling of constraints can drastically alter market clearing prices, reserve quantities, and revenue outcomes. Focusing on the diverse market designs and assumptions in ancillary services by U.S. RTOs, particularly in relation to capacity sharing and reserve substitutions, the research examines four distinct models that combine these elements based on a large-scale synthetic power system test data. Our study provides a critical insight into the economic implications and the underlying factors of these alternative reserve constraints through market simulations and data analysis.

## Nomenclature

*Indexes*:

| | |
|---|---|
| $g$ | Index of generators. |
| $t$ | Index of time intervals. |
| $l$ | Index of transmission lines. |
| $n$ | Index of buses. |
| $lc$ | Index of load curves. |
| $z$ | Index of reserve zones. |

*Parameters* (index $t$ denotes period):

| | |
|---|---|
| $\bar{F}_{l,t}$ | Maximum rate of line $k$. |
| $P_g^{min}$ | Minimum capacity of unit $g$. |
| $P_g^{max}$ | Maximum capacity of unit $g$. |
| $RSU_g$ | Start-up ramp rate of unit $g$. |
| $RSD_g$ | Shut-down ramp rate of unit $g$. |
| $RU_{g,t}^{5\,min}$ | 5 minutes ramp up rate of unit $g$. |
| $RU_{g,t}^{10\,min}$ | 10 minutes ramp up rate of unit $g$. |
| $RU_{g,t}^{60\,min}$ | 60 minutes ramp up rate of unit $g$. |
| $RD_{g,t}^{60\,min}$ | 60 minutes ramp down rate of unit $g$. |
| $RR_{t,z}^{REG}$ | Zonal regulation reserve requirements. |
| $RR_{t,z}^{SPIN}$ | Zonal spinning reserve requirements. |
| $RR_{t,z}^{NSP}$ | Zonal non-spinning reserve requirements. |
| $C_{g,t}^{SU}$ | Unit $g$ startup cost. |
| $C_{g,t}^{NL}$ | Unit $g$ no-load cost. |
| $C_{g,t}$ | Unit $g$ fuel cost. |
| $UT_g$ | Unit $g$ minimum up time. |
| $DT_g$ | Unit $g$ minimum down time. |
| $D_{n,t}$ | Electric demand at bus $n$. |
| $offer_{g,t}$ | Unit $g$ reserve offer. |
| $offer_{g,t}^{REG}$ | Unit $g$ regulation reserve offer. |
| $offer_{g,t}^{SPIN}$ | Unit $g$ spinning reserve offer. |
| $offer_{g,t}^{REG}$ | Unit $g$ non-spinning reserve offer. |
| $PTDF_{l,i}^{R}$ | Power Transfer Distribution Factors of line $l$ and bus $i$ based on reference bus $R$. |

*Variables* (index $t$ denotes period):

| | |
|---|---|
| $p_{g,t,lc}$ | Cleared energy on unit $g$ with load curve $lc$. |
| $p_{g,t}$ | Cleared energy on unit $g$. |
| $r_{g,t}^{REG}$ | Cleared regulation reserve on unit $g$. |
| $r_{g,t}^{SPIN}$ | Cleared spinning reserve on unit $g$. |
| $r_{g,t}^{NSP}$ | Cleared non-spinning reserve on unit $g$. |
| $u_{g,t}$ | Unit $g$ status. |
| $su_{g,t}$ | Startup variable of unit $g$. |
| $sd_{g,t}$ | Shutdown variable of unit $g$. |
| $f_{l,t}$ | Active power flow in line $l$. |

*Shadow prices* (indexes $t$ and $z$ denote period and zone, respectively):

| | |
|---|---|
| $SP_{t,z}^{REG}$ | First non-cascading constraint shadow price. |
| $SP_{t,z}^{SPIN}$ | Second non-cascading constraint shadow price. |
| $SP_{t,z}^{NSP}$ | Third non-cascading constraint shadow price. |
| $SP_{t,z}^{R}$ | First cascading constraint shadow price. |
| $SP_{t,z}^{RS}$ | Second cascading constraint shadow price. |
| $SP_{t,z}^{RSN}$ | Third cascading constraint shadow price. |

*Market clearing prices* (index $t$ denotes period):

| | |
|---|---|
| $MCP_{t,z}^{REG}$ | Regulation reserve MCP in zone z. |
| $MCP_{t,z}^{SPIN}$ | Spinning reserve MCP in zone z. |
| $MCP_{t,z}^{NSP}$ | Non-spinning reserve MCP in zone z. |

H. Davoudi, F. Wang, and D. Shi are with New Mexico State University, Las Cruces, NM, 88011 USA (e-mail: hdavoudi, fywang, dshi@nmsu.edu).

Y. Chen is with National Renewable Energy Lab, Golden, CO, 80401 USA (e-mail: Yonghong.chen@nrel.gov).

A. Xavier and F. Qiu are with Argonne National Lab, Lemont, IL, 60439 USA (email: axavier, fqiu@anl.gov)

$LMP_{n,t,lc}$   LMP at bus $n$ with load curve $lc$.

*Generators costs, revenues, and profits:*

$Ecos_{g,t,lc}$   Unit $g$ energy cost with load curve $lc$ at period $t$.
$Ecos_g$   Unit $g$ energy cost.
$Rcos_g^{onlineR}$   Unit $g$ online reserves cost.
$Rcos_g^{NSP}$   Unit $g$ non-spinning reserve cost.
$Erev_g$   Unit $g$ revenue from energy producing.
$Rrev_g^{REG}$   Unit $g$ revenue from regulation reserve clearing.
$Rrev_g^{SPIN}$   Unit $g$ revenue from spinning reserve clearing.
$Rrev_g^{NSP}$   Unit $g$ revenue from non-spinning reserve clearing.
$Epro_g$   Unit $g$ profit from energy producing.
$Rpro_g^{onlineR}$   Unit $g$ profit from online reserves clearing.
$Rpro_g^{NSP}$   Unit $g$ profit from non-spinning reserve clearing.
$LO_{g,t}$   Unit $g$ lost opportunity cost.
$unit\ cost_{g,t}^{onlineR}$   Unit $g$ lost opportunity cost to provide 1MW online reserves.

## I. Introduction

ANCILLARY service in electricity markets refers to a range of support services that are necessary to ensure the reliable operation of the electrical grid. These services are essential to maintain the balance between electricity supply and demand and to address various technical and operational challenges that arise in the power system. Ancillary services are typically procured and managed by grid operators or independent system operators (ISOs) to ensure grid stability and reliability. In electricity markets of United States ancillary service settlements consist of cost-based pricing and market-based services pricing. The pricing for cost-based products is generally fixed and regulated by the state or federal governments in the United States. However, market-based product prices are determined in a competitive electricity market with market clearing prices. Instead of depending on recurrent interactions between suppliers and consumers for establishing market equilibrium, a centralized electricity market introduces a structured mechanism to reach this equilibrium. Centralized electricity markets map the operational needs and requirements into mathematical constraints and provide market pricing that supports market equilibrium. Nevertheless, the modelling of generators' capacity and reserve requirement constraints could have a remarkable impact on the outcomes of the electricity market.

ISOs in the United States have different modeling and procedures, which will result in different market clearing, pricing, and settlement in ancillary service products. An energy and reserve co-optimization model, which takes lost opportunity cost and variable cost into account in ISO-New England market, is discussed in [1]. The primary conclusion of [1] is that, essentially, co-optimization of energy, reserve, and lost opportunity costs is a non-differentiable, potentially bilevel optimization problem. A simultaneous security-constrained market-clearing procedure is studied in [2], wherein the nodal marginal cost of security determines the price of reserve services. For primary, secondary, and tertiary services, it implements both probabilistic and deterministic models. In a short-term operation and pricing of energy and reserve, Arroyo et al. [3] take into account transmission line flow restrictions and preventive and corrective security. All ancillary products have a single price in the market clearing procedure provided by [2] and [3]. However, it does not seem to be satisfactorily accurate since different products impose diverse costs on the supplier. Thus, each product should have its own specific price. A co-optimized modeling of energy and reserves in nested zonal systems is presented in [4], taking into account the deliverability of reserves during first and second contingency occurrences. Network topology and transmission line limits could hinder the delivery of reserve. The authors of [4] acknowledge that because the zonal reserve model only takes the reserve zone interface limitation into account during reserve events, it is unable to address the issue of reserve deliverability adequately. To make sure the cleared reserve is deliverable, some operators manually disqualify the units having deliverability limits. In order to meet the requirements for zonal reserves while taking deliverability into account, Ref. [5] presents a co-optimized model that incorporates post-zonal reserve deployment transmission restrictions into the co-optimization of energy and ancillary services. It investigates how MCPs in zonal reserves are impacted by transmission limits. Nevertheless, the zonal approach cannot effectively evaluate the impact of reserve deployment on a transmission constraint within a single reserve zone. Hence, to address this issue, Ref. [6] introduced a nodal short-term reserve model for the Midcontinent Independent System Operator (MISO) in order to enhance post-event reserve deliverability and price signal. An approach for minimizing power transfers between zones so that both automatic and manual frequency restoration reserves contribute to meeting reserve needs is presented by Papavasiliou et al. [7]. A spinning reserve security constraint is provided in [8] for a post-contingency operation model that addresses state-of-charge deviation in energy storage systems following the deployment of spinning reserves. It is concluded that the above approach may result in a modest rise or decrease in individual participants' energy and reserve revenue due to changes in LMPs and MCPs. Still, it may also raise the overall system operating costs. Chen [9] enhances the designs of reserve products for MISO in order to enhance uncertainty management. A machine learning clustering technique defines seasonal hourly dynamic demand curves and market-wide requirements. A hot spinning reserve allocation is shown in [10] to address large unit failures during peak hours. To prevent cascading outages, load curtailment and effective reactive power allocation are implemented. The system configuration, cost coefficients of generators, and energy price of load curtailment all influence the optimal solution. Reference [11] analyses the impact of different ancillary services models with market-wide requirements on market outcomes considering zero offers for operating reserves.

Reserves force generators not to produce power in order to leave the headroom available to deploy reserves when needed. If a generator offers more than one reserve product type, the market outcomes will differ depending on whether the generation capacity is shared or not. This shows differing operating assumptions. Generators can supply various reserve products with the same headroom by sharing capacity. Sharing capacity is anticipated to lower the opportunity cost for generators to supply reserve products in order to meet reserve requirements. It is presumed that the same generation capacity can be cleared as several market products if capacity sharing amongst reserve products is permitted. Operators can represent the unit capacity constraints as sharing under this assumption. Reserve products do not share the generator's capacity in [4]-[5] and [8], but they do share it in [2]. However, [6] and [9] employ a mix of capacity constraints, which include sharing and non-sharing. The short-term reserve shares the capacity with the regulation, spinning, and supplemental reserves, yet these three do not share the capacity.

Furthermore, non-cascading reserve requirements refer to the fact that each sort of reserve requirement can be satisfied on its own. As an alternative, some reserves can support and substitute for others when requirements cascade. High-quality products can replace low-quality ones. Response time is the criterion used to determine the quality of reserves, the shorter the response time, the higher the quality. The non-cascading and cascading requirements are found in [2] and [4-5], respectively. Refs. [6-8] and [9] nevertheless model a combination of cascading and non-cascading requirements. Higher-quality products might substitute lower-quality products thanks to cascading reserve requirements. When a product satisfies more than one constraint at once, it demands different MCP calculations than when it is not cascading. Ancillary service providers bid strategically to maximize the profits by trading-off between the revenue and the possibility of not getting selected. Reference [12] presents a method to bid by a single Generating Company (GENCO) for real power and the ancillary services of reactive power and spinning reserve. Authors of [13] aim to maximize the revenue of microgrids containing wind turbines, photovoltaic systems, micro-turbines, and energy storage systems in providing energy, spinning reserve, and flexible ramping products (FRPs). They utilize hybrid stochastic/robust optimization in order to deal with uncertainties from renewable resources. Similarly, [14] and [15] propose a stochastic bidding strategy to help a wind power plant and pumped-storage plant in reducing imbalances in bids and penalties in the day-ahead market caused by uncertainties.

TABLE I
ALLOWANCE OF NON-ZERO OFFERS IN ANCILLARY SERVICE PRODUCTS [16]

|  | ISO-NE | NYISO | PJM | MISO | SPP | ERCOT | CAISO |
|---|---|---|---|---|---|---|---|
| Regulation | Yes | Yes | Yes | Yes | Yes | Yes | Yes |
| Secondary Contingency | No for TMSR, TMNS, TMOR | Yes in day-ahead, No in real-time | SR-Yes NSR - No | Yes | Yes | Yes | Yes |

Acronyms: TMSR: ten-minute spinning reserve; TMNSR: ten-minute non-spinning reserve; TMOR: Thirty-minute operating reserve; SR: synchronized reserve; NSR: non-synchronized reserve.

However, in some ISOs, non-zero offers are not allowed for some products [16]. The allowance of non-zero offers in ancillary service products is presented in Table I. Market participants bid for regulation reserve in all ISOs. Nevertheless, non-zero offers for ten-minute spinning, non-spinning, and thirty-minute operating reserves are not allowed in ISO-NE. Similarly, market participants do not bid for secondary contingency products in New York Independent System Operator (NYISO) real-time market and for non-synchronized reserve in Pennsylvania-New Jersey-Maryland Interconnection (PJM).

The marginal cost to provide the next increment (MWh) of the service is the Market Clearing Price (MCP). MCPs for products with zero offers are calculated as lost opportunity cost, which implies the lost profit by withholding the capacity not producing energy. FERC Order 755 (2011) required all FERC-jurisdictional ISOs to include lost opportunity costs as a pricing mechanism for regulating reserve. The market participants providing regulating reserve would get paid the lost opportunity cost of the marginal unit [17].

The contributions of this paper are as follows,
1) This paper investigates different combinations of capacity sharing and reserve substitutions among different reserve products, which are not consistent in U.S. RTOs. The modeling and market implications of capacity sharing and reserve substitutions are not thoroughly examined in the existing literature. However, slight differences in market modeling may lead to significant market outcomes.
2) This paper quantifies the resulting impacts on market outcomes, including LMPs, MCPs, cleared reserve quantity, generation system-wide/individual cost, revenue, and profit as well as objective function.
3) This paper investigates how the profitability of various power plants (based on fuel type) are affected by different market designs and provides the underlying factors contributing to the resulting market outcomes.

The remainder of this paper is organized as follows. Section II discusses electricity market modeling with reserve constraint and pricing. Numerical results are shown in Section III. Section IV concludes this paper and offers some future research.

## II. ELECTRICITY MARKET MODELING

The active power capacity above or below the scheduled energy to be deployed in case of a contingency or load variation in order to maintain active power balance is known as operating reserve products. Three categories of reserves are taken into consideration in this study:
1) Regulation reserve (REG): reduce frequency deviation and interchange errors by minimizing the Area Control Error (ACE). The regulation reserve deployment is managed by Automatic Generation Control (AGC). It is a non-event product that reacts to variations in demand, including unexpected demand spikes and variable energy resource variances continually.
2) Spinning reserve (SPIN): an event reserve option for unforeseen transmission or generation failures. A synchronized unit can typically supply the spinning reserve if it can ramp up to the dispatch target of the balancing authority in less than ten minutes and continue to operate for several hours prior to the reserve capacity being replenished.
3) Non-spinning reserve (NSP) for post-event system support is provided by fast-start offline units.

It may be possible to share capacity and use different reserve products as reserve substitutes due to variations in the deployment logics and response times of operating reserves. The purpose of this study is to examine various combinations of reserve substitutes and capacity sharing in day-ahead markets and to comprehend the resulting effects on the market.

### A. Generators Capacity Constraints

Reserves take advantage of the dispatchable backup capacity (headroom for upward and rear room for downward reserve). The capacity sharing of energy, regulation reserve, and spinning reserve is examined in this paper. The constraints (1a) and (1b) denote the non-sharing capacity constraints, meaning that the units' capacity is not shared by regulation and spinning reserves (option 1). However, in option 2, the capacity can be employed concurrently to clear both using the capacity sharing constraints (2a)–(2d).

Option 1: non-sharing generation capacity

$$p_{g,t} - r_{g,t}^{REG} - r_{g,t}^{SPIN} \geq u_{g,t} P_g^{min} \quad (1a)$$
$$p_{g,t} + r_{g,t}^{REG} + r_{g,t}^{SPIN} \leq u_{g,t} P_g^{max} \quad (1b)$$

Option 2: sharing generation capacity

$$p_{g,t} - r_{g,t}^{REG} \geq u_{g,t} P_g^{min} \quad (2a)$$
$$p_{g,t} + r_{g,t}^{REG} \leq u_{g,t} P_g^{max} \quad (2b)$$
$$p_{g,t} - r_{g,t}^{SPIN} \geq u_{g,t} P_g^{min} \quad (2c)$$
$$p_{g,t} + r_{g,t}^{SPIN} \leq u_{g,t} P_g^{max} \quad (2d)$$

When a unit becomes online from offline status (start-up process), it can produce at most as its start-up ramp rate. This physical limitation is modeled by start-up constraints, which change based on capacity constraints (above options). In the non-sharing capacity option, the regulation and spinning reserves do not share the start-up ramp rate (3). However, the regulation and spinning reserves are allowed to occupy the start-up ramp rate in the meantime, in sharing capacity option (4a)-(4b). The term $(1 - su_{g,t})P_g^{max}$ restricts the sum of energy and reserves by the maximum capacity in the other intervals. For the units with a lower ramp rate than the minimum capacity (e.g., nuclear units), the start-up ramp rate can be set as the technical minimum avoiding infeasibility.

Start-up constraint in non-sharing capacity:
$$p_{g,t} + r_{g,t}^{REG} + r_{g,t}^{SPIN} \leq su_{g,t} RSU_g + (1 - su_{g,t})P_g^{max} \quad (3)$$

Start-up constraint in sharing capacity:
$$p_{g,t} + r_{g,t}^{REG} \leq su_{g,t} RSU_g + (1 - su_{g,t})P_g^{max} \quad (4a)$$
$$p_{g,t} + r_{g,t}^{SPIN} \leq su_{g,t} RSU_g + (1 - su_{g,t})P_g^{max} \quad (4b)$$

### B. Zonal Reserve Requirements Constraints

Lower quality service products can be replaced by reserves with faster response times, which are expected to offer a higher level of service. To be more precise, spinning reserve and non-spinning reserve can be substituted with regulation reserve; similarly, non-spinning reserve can be substituted with spinning reserve. Cascading reserve requirements are the name given to these reserve requirement constraints. It should be noted that, in the case that reserve substitution is permitted, sharing generation capacity amongst reserve products may result in double pay generation units. As option 1, non-cascading regulation, spinning, and non-spinning reserve requirements are the constraints (5a) - (5c), in that order. Every type of reserve satisfies its own zonal requirements. However, given the cascade constraints (6a)–(6c), the regulation reserve could meet both the spinning and non-spinning zonal requirements, in addition to meeting its own zonal requirements. Likewise, in addition to meeting its zonal requirements, the spinning reserve may also fulfil non-spinning requirements (option 2). For every constraint, the corresponding shadow prices are shown by the symbols in brackets.

Option 1: non-cascading
$$\sum_{g \in G_z} r_{g,t}^{REG} \geq RR_{t,z}^{REG}, \{SP_{t,z}^{REG}\} \quad (5a)$$
$$\sum_{g \in G_z} r_{g,t}^{SPIN} \geq RR_{t,z}^{SPIN}, \{SP_{t,z}^{SPIN}\} \quad (5b)$$
$$\sum_{g \in G_z} r_{g,t}^{NSP} \geq RR_{t,z}^{NSP}, \{SP_{t,z}^{NSP}\} \quad (5c)$$

Option 2: cascading
$$\sum_{g \in G_z} r_{g,t}^{REG} \geq RR_{t,z}^{REG}, \{SP_{t,z}^{R}\} \quad (6a)$$
$$\sum_{g \in G_z} (r_{g,t}^{REG} + r_{g,t}^{SPIN}) \geq RR_{t,z}^{REG} + RR_{t,z}^{SPIN}, \{SP_{t,z}^{RS}\} \quad (6b)$$
$$\sum_{g \in G_z} (r_{g,t}^{REG} + r_{g,t}^{SPIN} + r_{g,t}^{NSP}) \geq RR_{t,z}^{REG} + RR_{t,z}^{SPIN} + RR_{t,z}^{NSP},$$
$$\{SP_{t,z}^{RSN}\} \quad (6c)$$

All reserve requirement constraints are modeled as hard constraints, indicating that reserve requirements must be satisfied without exception. Alternatively, these constraints could be modeled as soft constraints, where violations are allowed by penalizing the total cost. In such a design, shadow prices, and consequently MCPs, are capped and cannot exceed a certain amount. However, we prefer to use hard constraints to have unrestricted MCPs and investigate the influence of different models on market outcomes precisely.

### C. Market Clearing Prices, Cost, Revenue, and Profit

The shadow price of a zonal requirements constraint is the MCP for each type of reserve in that zone in non-cascading reserve requirements (7), (8), and (9). On the other hand, zonal MCPs for regulation, spinning, and non-spinning reserves in cascade requirements are determined by (10), (11), and (12), respectively. Since the regulation reserve meets all three cascade constraints, the associated MCPs are the total of their shadow prices. Furthermore, because the spinning reserve satisfies (6b) and (6c) constraints, corresponding MCPS are the total of the shadow prices for those two constraints.

$$MCP_{t,z}^{REG} = SP_{t,z}^{REG} \quad (7)$$
$$MCP_{t,z}^{SPIN} = SP_{t,z}^{SPIN} \quad (8)$$
$$MCP_{t,z}^{NSP} = SP_{t,z}^{NSP} \quad (9)$$

$$MCP_{t,z}^{REG} = SP_{t,z}^{R} + SP_{t,z}^{RS} + SP_{t,z}^{RSN} \quad (10)$$
$$MCP_{t,z}^{SPIN} = SP_{t,z}^{RS} + SP_{t,z}^{RSN} \quad (11)$$
$$MCP_{t,z}^{NSP} = SP_{t,z}^{RSN} \quad (12)$$

Observe that $SP_{t,z}^{R}, SP_{t,z}^{RS}, SP_{t,z}^{RSN} \geq 0$ based on duality. As a result, it is ensured that higher-quality services will get zonal market clearing prices no less than lower-quality services in the case of cascading reserve requirements. In this study $MCP_{t,z}^{REG} \geq MCP_{t,z}^{SPIN} \geq MCP_{t,z}^{NSP}$.

The energy-producing generators' cost, revenue, and profit are computed by (13), (14), and (15), respectively. Energy cost is calculated based on the fuel cost of units. The shadow price of the nodal power balancing constraint (33) at bus $n$ and interval $t$ is denoted by $LMP_{n,t}$.

$$Ecos_g = \sum_t C_{g,t} p_{g,t} \quad (13)$$
$$Erev_g = \sum_t LMP_{n,t} p_{g,t} \quad (\textit{g is located in bus } n) \quad (14)$$

$$Epro_g = Erev_g - Ecos_g \tag{15}$$

As already mentioned, market participants imposed lost opportunity cost to provide reserve products, which is the profit of the producing energy instead of the providing reserve. Lost opportunity cost is calculated by (16). In some cases (as an instance when a unit is producing at its minimum capacity limit), it is possible that the LMP is less than the production cost, which means the unit is making a negative profit $LO_{g,t} < 0$. The unit reserve cost is considered zero in these cases (17). However, when $LO_{g,t} \geq 0$, it is considered as $LO_{g,t}$ (18). This unit cost is for products provided by synchronized units, which are regulation and spinning reserves in this study. We will call them online reserves.

$$LO_{g,t} = LMP_{n,t} - \frac{Ecost_{g,t}}{p_{g,t}}, \text{ (g is located in bus n)} \tag{16}$$

$$\text{If } LO_{g,t} < 0 \quad unit\ cost_{g,t}^{onlineR} = 0 \tag{17}$$

$$\text{If } LO_{g,t} \geq 0 \quad unit\ cost_{g,t}^{onlineR} = LO_{g,t} \tag{18}$$

Units' capacity is used as the summation of the online reserve in non-sharing models; thus, the online reserve cost is calculated by (19) in non-sharing cases. Nevertheless, the capacity is occupied as the larger amount between the cleared regulation and spinning products. As a result (20) is used to calculate the online reserves cost in sharing capacity models.

$$Rcos_g^{onlineR} = unit\ cost_{g,t}^{onlineR} \left( r_{g,t}^{REG} + r_{g,t}^{SPIN} \right) \tag{19}$$

$$Rcos_g^{onlineR} = unit\ cost_{g,t}^{onlineR}\ maximum \left( r_{g,t}^{REG}, r_{g,t}^{SPIN} \right) \tag{20}$$

The offer of non-spinning reserve is its unit cost, and the total cost is calculated as (21).

$$Rcos_g^{NSP} = \sum_t offer_{g,t}^{NSP}\ r_{g,t}^{NSP} \tag{21}$$

Generators' revenue from offering regulation, spinning, and non-spinning reserves are calculated by (22a)-(22c), respectively. Each zone has its own MCP, and to calculate a unit revenue, the MCP of the corresponding zone is used.

$$Rrev_g^{REG} = \sum_t MCP_{t,z}^{REG} r_{g,t}^{REG} \quad , g \in Gz \tag{22a}$$

$$Rrev_g^{SPIN} = \sum_t MCP_{t,z}^{SPIN} r_{g,t}^{SPIN} \quad , g \in Gz \tag{22b}$$

$$Rrev_g^{NSP} = \sum_t MCP_{t,z}^{NSP} r_{g,t}^{NSP} \quad , g \in Gz \tag{22c}$$

Equations (23a)-(23b) are to calculate units' profit from offering online, and non-spinning, respectively.

$$Rpro_g^{onlineR} = (Rrev_g^{REG} + Rrev_g^{SPIN}) - Rcos_g^{onlineR} \tag{23a}$$

$$Rpro_g^{NSP} = Rrev_g^{NSP} - Rcos_g^{NSP} \tag{23b}$$

### D. Objective Function

In this study, load and reserve requirements are considered as perfectly inelastic goods. It means that all load and ancillary services must be supplied at any cost. In such a market, the consumers' surplus is zero, and the producers' surplus and social welfare are equal. As producers' surplus is equal to producers' revenue minus producers' cost, the smaller cost will cause the larger surplus. Consequently, minimizing total cost is equivalent to maximizing social welfare. The objective function is shown in (24), which includes startup, no-load, fuel, and reserve cost for each ancillary service product.

$$obj = min \sum_{g \in G, t \in T} (C_{g,t}^{SU} su_{g,t} + C_{g,t}^{NL} u_{g,t} + C_{g,t} p_{g,t} + offer_{g,t}^{REG} r_{g,t}^{REG} + offer_{g,t}^{SPIN} r_{g,t}^{SPIN} + offer_{g,t}^{NSP} r_{g,t}^{NSP}) \tag{24}$$

### E. Other Unit Commitment Constraints

Inequalities (25) and (26) represent the energy ramp-up and down constraints, respectively. Regulation and spinning reserves are constrained by 5- and 10-minute ramp rates, respectively, according to (27)–(28). Energy, regulation and spinning reserves share the ramp rate. Eligible offline generators are required by inequality (29) to provide the non-spinning reserve at a level lower than their start-up ramp limit.

$$p_{g,t} - p_{g,t-1} \leq su_{g,t} RSU_g + (1 - su_{g,t}) RU_{g,t}^{60\ min} \tag{25}$$

$$p_{g,t-1} - p_{g,t} \leq sd_{g,t} RSD_g + (1 - sd_{g,t}) RD_{g,t}^{60\ min} \tag{26}$$

$$r_{g,t}^{REG} \leq RU_{g,t}^{5\ min} \tag{27}$$

$$r_{g,t}^{SPIN} \leq RU_{g,t}^{10\ min} \tag{28}$$

$$r_{g,t}^{NSP} \leq (1 - u_{g,t}) RSU_g \tag{29}$$

Once a unit is started up, it must stay online at least for its minimum up time, and then it can be shut down, which is modeled by (30). Similarly, once a unit is shut down, it must stay offline for at least its minimum downtime, and then it can be started up (31).

$$\sum_{q=t-UT_g+1}^{t} su_{g,q} \leq u_{g,t} \tag{30}$$

$$\sum_{q=t-DT_g+1}^{t} sd_{g,q} \leq 1 - u_{g,t} \tag{31}$$

Constraint (32) represents the following: 1) When a unit is online while it was offline in the previous interval, the startup and shutdown variables for the current interval must be 1 and 0, respectively. 2) When a unit is offline while it was online in the previous interval, the startup and shutdown variables for the current interval must be 0 and 1, respectively. 3) When there is no status change between the previous and current interval, both startup and shutdown variables for the current interval must be 0. 4) Both startup and shutdown variables cannot be 1 in the same interval.

$$u_{g,t} - u_{g,t-1} = su_{g,t} - sd_{g,t} \tag{32}$$

The summation of power flow into and from a bus, generation, and demand at that bus shall be zero, which is called the nodal power balance (33). The shadow price of this constraint is the LMP for the corresponding bus.

$$\sum_{\forall l \in \delta(n)^+} f_{l,t} - \sum_{\forall l \in \delta(n)^-} f_{l,t} + \sum_{\forall g \in g(n)} p_{g,t} = D_{n,t}, \{LMP_{n,t}\} \tag{33}$$

The N-1 security constraints for transmission lines are modeled with PTDFs and nodal injections in equations (34) and (35). To be consistent with RTOs' practice, we only model a subset of the security constraints that are likely to bind to maintain the economic dispatch model computationally manageable.

$$f_{l,t} = PTDF_{n,l,t} \sum_{n \in N} (\sum_{g \in G(n)} p_{g,t} - D_{n,t}) \tag{34}$$

$$-\bar{F}_{l,t} \leq f_{l,t} \leq \bar{F}_{l,t} \tag{35}$$

## III. NUMERICAL RESULTS

Based on an open-source unit commitment (Unitcommitment.jl) [18], the optimization unit commitment problem and electricity market have been solved for the Electric Reliability Council of Texas (ERCOT) system [19], in four models:

- Model 1: non-sharing generator capacity and non-cascading reserve requirements. (NS-NC)
- Model 2: non-sharing generator capacity and cascading reserve requirements. (NS-C)
- Model 3: sharing generator capacity and non-cascading reserve requirements. (S-NC)
- Model 4: sharing generator capacity and cascading reserve requirements. (S-C)

Since most of the ISOs adopt non-zero offers for operating reserve, this paper assumes non-zero offers for the simulation studies. The procedure for offers calculation is illustrated in Fig. 1. These operating reserve offers are considered as the expected lost opportunity cost based on the average profits, which capture market dynamics with different seasonal, weekdays, weekend load patterns. When LMP is less than the average production cost, the calculated reserve cost in step 2 is negative. In other words, the unit wishes to provide reserve even for free instead of generating energy for negative profit. As a result, in such cases, reserve offers are set to zero in step 4 (38). Reserve offers are equal to the lost opportunity cost for the case with a positive reserve cost (39). Sometimes LMPs become negative due to congestion or subsidies such as renewable energy credits. This may result in a negative calculated reserve cost in step 2 and impacts the average lost opportunity cost.

The calculated offer is considered for the spinning reserve. The regulation and non-spinning offers are computed by multiplying spinning offers into 3.28 and 0.0864, respectively. These coefficients are derived from MISO day-ahead regulation products averaged price [16]. The proposed method for calculating reserve offers ensures that no regulation or spinning offers are lower than the profit made by producing energy instead of providing reserves. Additionally, this market operates under a uniform pricing mechanism (not a pay-as-bid pricing market), meaning participants are paid based on the marginal unit offers. As a result, participants can potentially earn even higher profits. This market design inherently incentivizes participants to provide reserve products.

Regulation reserve generally requires twice ramp rate than spinning. Furthermore, FERC Order 755 requires that all ISOs pay all regulating resources both a capacity payment and a performance payment (also called movement or mileage payment) [17]. Performance is measured as the actual upward or downward movement of a regulating resource as directed by the AGC.

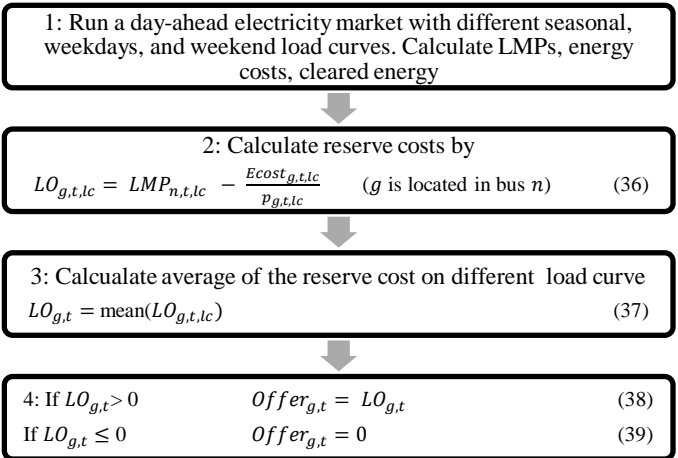

Fig. 1. Offers calculation procedure

The ERCOT system is a synthetic, large-scale system with 6717 buses, 9140 branches, and 731 generators [19]. Such a large system should be partitioned into reserve zones to avoid reserve deliverability issues due to transmission line congestion. The ERCOT system is partitioned into six zones using the method presented by [20-22]. This method utilizes K-means to cluster buses with respect to the Power Transfer Distribution Factors (PTDF). The PTDF of two buses are considered as the criteria of their impact on transmission lines; the closer PTDFs are to each other, the more similar their impact on transmission lines.

TABLE II
ZONING RESULTS

| | Z1 | Z2 | Z3 | Z4 | Z5 | Z6 |
|---|---|---|---|---|---|---|
| **Buses Number** | 1,914 | 1,597 | 647 | 1,127 | 1,264 | 168 |
| **Generators Number** | 112 | 184 | 89 | 176 | 140 | 30 |
| $GC_z^*$ **(MW)** | 21,426 | 25,556 | 8,880 | 26,500 | 17,359 | 5,190 |
| **Peak Load (MW)** | 25,076 | 20,634 | 5,721 | 9,409 | 1,2400 | 1,426 |
| $RR_{t,z}^{REG}$**(MW)**** | 604 | 497 | 138 | 227 | 299 | 34 |
| $RR_{t,z}^{REG}/GC_z^*$**(%)** | 2.82 | 1.94 | 1.55 | 0.85 | 1.72 | 0.66 |
| $offer_{gz}^{REG}$ **($/MWh)**** | 73.8 | 79.1 | 94.3 | 99.3 | 71.9 | 120.2 |
| $RR_{t,z}^{SPIN}$**(MW)**** | 1,205 | 1,280 | 785 | 932 | 655 | 329 |
| $RR_{t,z}^{SPIN}/GC_z^*$**(%)** | 5.62 | 5.01 | 8.84 | 3.52 | 3.77 | 6.34 |
| $offer_{gz}^{SPIN}$ **($/MWh)**** | 22.5 | 19.8 | 29.1 | 30.3 | 21.8 | 36.6 |
| $RR_{t,z}^{NSP}$**(MW)**** | 602 | 640 | 392 | 466 | 327 | 164 |
| $RR_{t,z}^{NSP}/GC_z^*$**(%)** | 2.81 | 2.50 | 4.42 | 1.76 | 1.89 | 3.17 |
| $offer_{gz}^{NSP}$ **($/MWh)**** | 6.4 | 5.6 | 8.3 | 8.6 | 6.2 | 10.4 |

* Generation capacity in zone $z$
** Values are average.

Partitioning results are presented in Table II. Like the Pennsylvania, New Jersey, and Maryland (PJM) RTO reserve requirements, spinning and non-spinning zonal reserve requirements are determined 100 percent and 50 percent of the largest single contingency in that zone, respectively [16]. This ensures that the system is able to meet N-1 criteria on the generation side. The regulation reserves requirement is modeled as a fraction of the hourly load, which reflects the real-time load deviation. Each zone contains a diverse amount of generation capacity ($GC_z^*$); for example, Z4 has the highest (26,500 MW); thus, the proportion of the regulation, spinning, and non-spinning requirements to generation capacity (0.85, 3.52, and 1.76 %) are relatively low. Moreover, the average offers of each product are calculated in each zone; for instance, the average of the regulation, spinning, and non-spinning reserves in Z4 are 99.3, 30.3, and 8.6 $/MWh, respectively. It is noteworthy that although zone six has the lowest generation capacity, it has the highest offers for all ancillary services and relatively high proportions of spinning and non-spinning requirements.

*A. Market Clearing Prices (MCPs) Analysis*

MCPs will be discussed in three aspects: 1) comparing different products MCPs (Fig. 2 a). 2) analyzing the impact of the models on MCPs (Fig. 3 a-c). 3) Studying zoning influence on MCPs (Fig. 3 d-f). Fig. 2 a shows regulation, spinning, and non-spinning MCPs in the NS-NC model and zone six. As mentioned, regulation reserve requires twice the ramp rate than spinning, and regulation offers are usually greater than spinning and non-spinning, and spinning offers are usually greater than non-spinning offers.

Sharing capacity between regulation and spinning products significantly affects MCPs. Due to more available and cheaper capacity in sharing models for regulation and spinning reserves, the associated MCPs are considerably cheaper in S-NC and S-C (Fig. 3 a and b). However, in non-sharing models, the headroom of the cheaper units can be used only once to clear either regulation or spinning reserve. The other one should be cleared from relatively more expensive units. Therefore, online reserve MCPs are more expensive in NS-NC and NS-C (Fig. 3 a and b). On the other hand, cascading requirements bring a more flexible and extended feasible region for the market optimization model, and it is expected to reduce the total cost (objective function) when substitution happens. There are different factors in the objective function (24) to minimize total cost, and generators' startup cost is one of them. In cascading models, the solver clears reserves from slightly expensive units and commits fewer units rather than committing more units and imposing more startup costs to clear reserves from slightly cheaper units. Therefore, cascading models have less startup cost (Fig. 2. b) and, consequently, less objective function (Fig. 2. c) compared to models with the same capacity constraints.

Zoning impact on MCPs is indicated in Fig. 3 d-f. Supply quantity in each zone, which is generation capacity, and units' offers play an important role in MCPs calculation. Zones with small generation capacity because of capacity scarcity, and high offers lead to more expansive MCPs. Zone six has the greatest average offers for all products. Having the smallest generation capacity, it has the second highest proportion of spinning and non-spinning requirements. Therefore, zone six has the most expensive MCPs. Zone three has the largest proportion of spinning and non-spinning requirements, and the third highest offers for all products. As a result, it has the second most expensive MCPs in most of the intervals. Although the second highest offers for all products are in zone four, it has the greatest generation capacity, consequently, the smallest proportion of spinning and non-spinning requirements and the second smallest proportion of regulation. The abundance of supply reduces the effect of high offers and makes zone four MCPs moderate. Zones one, two, and five have relatively close offers and generation capacity. Thus, MCPs are tight in these zones. In the intervals with zero MCPs there are enough units with zero offers to fulfill the hourly corresponding requirements, which happens in zones with a large generation capacity and a relatively small proportion of requirements, such as zones one, two, four, and five.

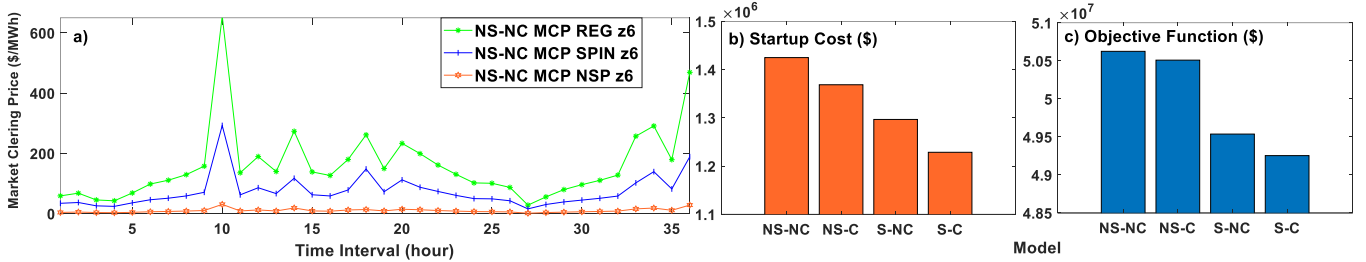

Fig. 2. a) NS-NC MCPs in zone six. b) Startup costs for ERCOT system. c) Objective functions

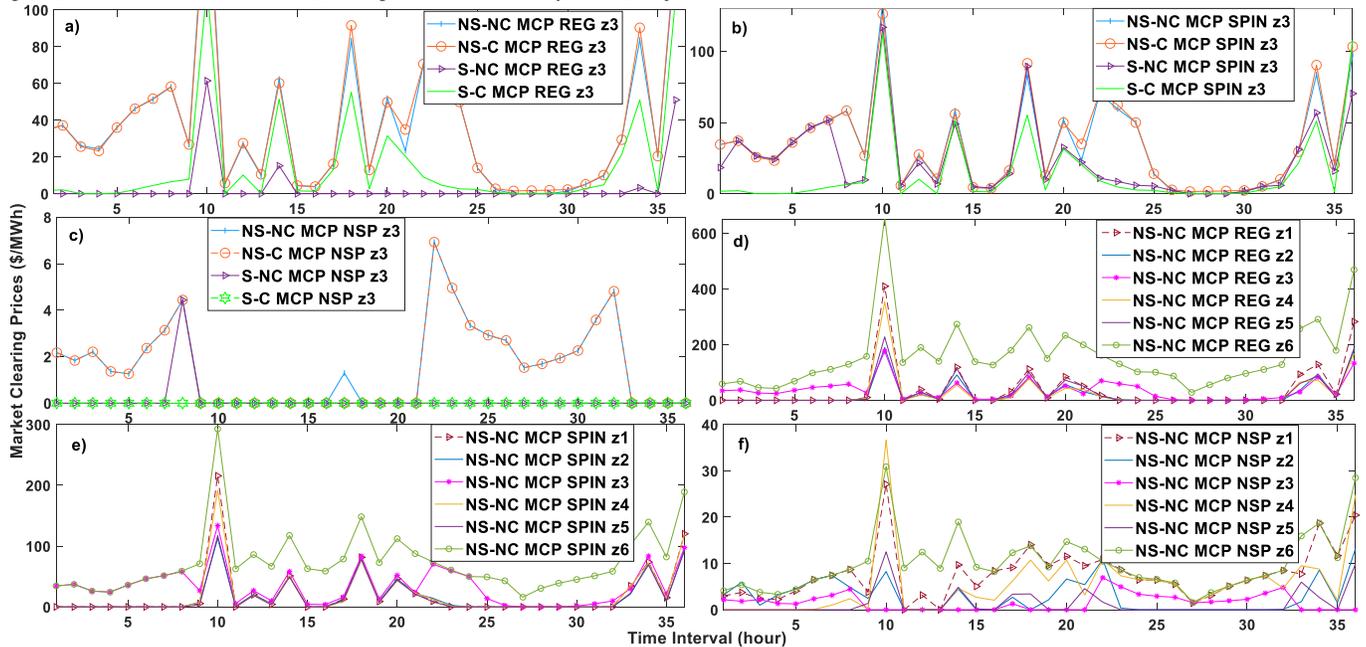

Fig. 3. a, b, and c) regulation, spinning, and non-spinning MCPs in zone 3 for all models, respectively. d, e, and f) regulation, spinning, and non-spinning MCPs for NS-NC in all zones, respectively

### B. Revenue, Cost, and Profit Analysis

The generator's revenue, cost, and profit from producing energy and clearing each ancillary service are presented in Fig. 4. Available capacity is lower in models with non-sharing capacity versus models with sharing capacity constraints. Therefore, units with more expensive fuel costs and reserve offers are utilized to fulfill the load and reserve requirements. It means the marginal units have higher fuel costs and reserve offers, which make higher LMPs and MCPs. As a result, NS-NC has considerably greater revenue, cost, and profit for all products compared to S-NC, as well as NS-C compared to S-C. Higher-quality products can be substituted with lower-quality products in cascading reserve requirements. The cascading models reduce the total cost by committing fewer units and lowering startup costs. Therefore, NS-C has lower total cost than NS-NC, as well as S-C than S-NC (Fig. 4 h). However, committing fewer units may make LMPs and online reserves MCPs slightly higher in some intervals. Consequently, energy and online reserves' revenue, cost, and profit are slightly larger in cascading models compared to non-cascading. Nevertheless, as many online units can provide cheaper regulation and spinning reserves to substitute the non-spinning, non-spinning MCPs are smaller in cascading models. Therefore, NS-C has lower non-spinning revenue, cost, and profit versus NS-NC, as well as S-C versus S-NC (Fig. 4 c, g, and k).

As shown in Fig. 4 f, online reserves cost in sharing capacity models is noticeably less than in non-sharing models. The reason is that the summation of the cleared regulation and spinning reserve is used for online reserve cost calculation in non-sharing models (19). Yet only the greater amount between cleared regulation and spinning is used in sharing models (20). In other words, the overlapping capacity between regulation and spinning reserve is considered only once in cost computing. Thus, it leads to a much smaller online reserve cost.

As elucidated earlier, sharing capacity and cascading reserve requirements reduce the total cost. Based on Fig. 4 h, sharing capacity has more significant impact in comparison to cascading requirements because the cost difference between NS-NC and S-NC is much more than the cost different between NS-NC and NS-C. S-C has the smallest total cost because it is modeled with both sharing capacity and cascading requirements. Since the highest LMPs and MCPs caused by non-sharing capacity and cascading requirements constraints, NS-C has the highest total revenue and profit (Fig. 4 d and I). NS-NS has the second-greatest total revenue and profit due to the second-highest energy and online reserves and the highest non-spinning revenue. Sharing capacity is more effective than non-cascading requirements in decreasing LMPs and MCPs. Therefore, S-NC has less total revenue and profit than NS-NC.

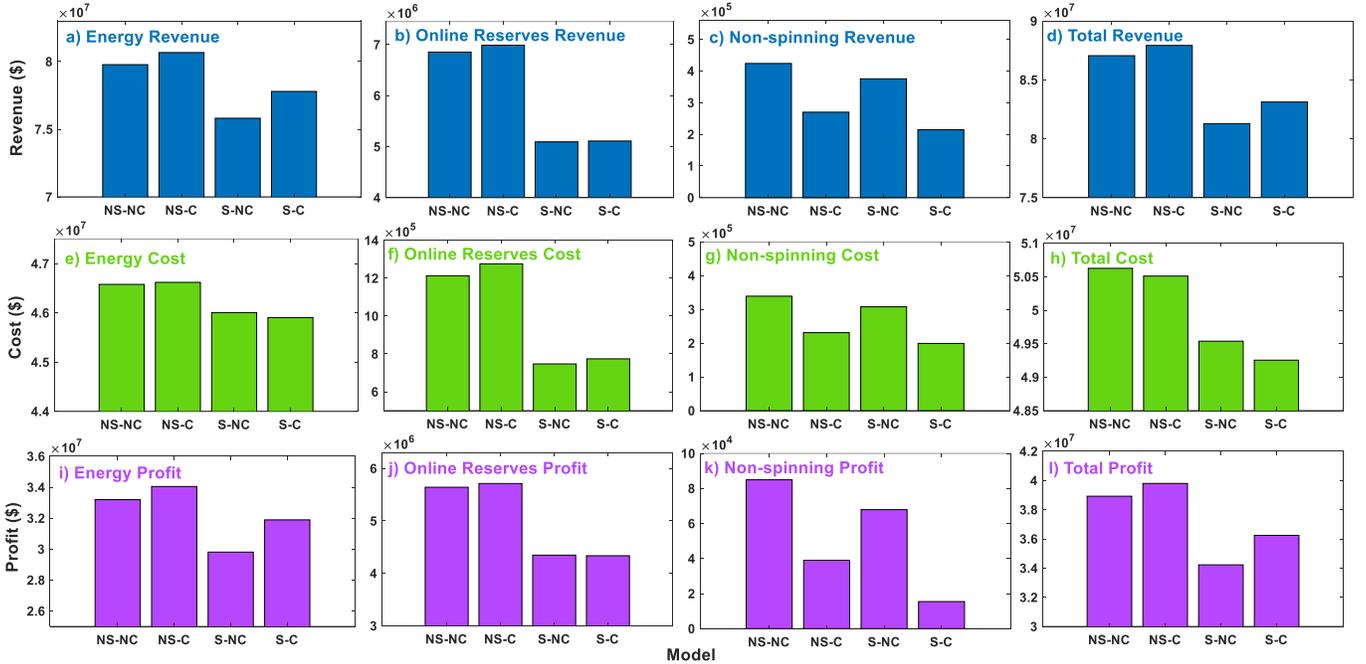

Fig. 4. Revenue, cost, and profit for energy and ancillary services

The synthetic Texas system's units are categorized based on fuel type to study the influence of the models on different types of power plants, and results are indicated in Table III. Natural gas power plants (NG) with the highest share cover more than half of the total generation capacity. Wind (WND) and bituminous coal (BIT) units have the second and third greatest share, with 24.09% and 13.73%, respectively. 4.73% and 2.23% of the total capacity are supplied by nuclear (NUC) and solar (SOL) power plants, respectively. Hydro (HYD), and other (OTH), covers less than 1%. We assume that all forecasted capacities for renewable units are firm and those are equipped with grid-forming inverters and capable of supplying reserves. Even though forecasts of WND and SOL are uncertain, the real-time market will settle for difference.

The average of the fuel costs in the up limit, startup cost, ramp rate, and offers for regulation, spinning, and non-spinning reserves are shown in Table III. The fuel costs are zero for renewable units (SOL, WND, and HYD), meaning their energy production profit is equal to their revenue. Therefore, they incur a significant lost opportunity cost to provide reserves and have the highest offers for ancillary services. NG has expensive fuel costs, which make the energy profit, and consequently, the lost opportunity cost lower. Thus, it has lower ancillary services offers. Moreover, the high ramp rate (764.5 MW/h) makes NG capable of providing online reserves, especially regulation and following the load.

On the other hand, BIT and NUC have the cheapest fuel cost after renewable, substantially high startup costs, moderate reserve offers, and a very long response time. Therefore, they are unsuitable for providing reserves because it requires a relatively short response time. OTH is not analyzed since it includes different types of units with different characteristics, such as energy storage, petroleum coke, hybrid renewable, etc., and covers a miner portion of the capacity (0.86%).

TABLE III
UNITS CATEGORIZATION BASED ON FUEL TYPE

|  | NG | BIT | NUC | SOL | WND | HYD | OTH |
|---|---|---|---|---|---|---|---|
| Number | 475 | 23 | 4 | 36 | 153 | 22 | 18 |
| $GC^*$ (%) | 53.89 | 13.73 | 4.73 | 2.23 | 24.09 | 0.47 | 0.86 |
| Average $C_g^{SU}$ ($) | 2959 | 22041 | 29097 | 0 | 0 | 0 | - |
| Average $C_g$ ($) | 49.9 | 24 | 12.9 | 0 | 0 | 0 | - |
| Average $RU_g^{60\ min}$ (MW/h) | 764.5 | 60 | 6 | UN** | UN** | UN** | - |
| Average $offer_g^{REG}$ ($/MWh) | 64.85 | 41.51 | 79.98 | 121.48 | 122.22 | 121.42 | - |
| Average $offer_g^{SPIN}$ ($/MWh) | 19.75 | 12.64 | 24.35 | 36.99 | 37.22 | 36.97 | - |
| Average $offer_g^{NSP}$ ($/MWh) | 5.60 | 3.59 | 6.91 | 10.50 | 10.56 | 10.49 | - |

* Generation capacity of each type
** Unlimited

Share of each type of unit in energy, ancillary services, and total revenue are presented in Fig. 5. Based on Fig. 5 a, since zero fuel and startup costs for SOL, WND, and HYD, they are used for producing energy as much as possible, unless transmission line congestion does not allow to clear energy from them. Similarly, BIT generates energy with the majority of the capacity because it has relatively cheap fuel costs, and only a slight amount of its capacity is scheduled for ancillary services due to low ramp rates and moderate reserve offers. Almost the entire capacity of NUC is utilized to produce energy due to the cheapest fuel cost. Additionally, the minor ramp rate does not allow NUC to change the initial output considerably and makes it unable to provide reserves. As a result, any reserve products are not cleared from NUC. As NG has a much more share in system capacity compared to others and a high ramp rate, almost half of the energy revenue is made by NG. Different models seem to not have a remarkable impact on energy revenue distribution.

According to Fig. 5 b, more than 84% of all ancillary revenue is earned by NG due to its high ramp rate and plenty of capacity. WND has the greatest share after NG in reserves providing. Although it has expensive reserve offers (almost three times as much as BIT and twice as much as NG), it has an extremely high ramp rate to respond to load deviation quickly. WND covers more than 10% of total revenue, although it only supplies approximately 1% of reserves. BIT and OTH supply a tiny amount of ancillary services. Ancillary service revenue for the rest of the units is ignorable. The main competition in the reserves revenue is between NG and WND because: 1) they together have 88% of the total capacity, 2) they have a high ramp rate, and 3) other units either cover minor capacity or are very slow to provide reserve products. Sharing capacity affects the distribution of the ancillary services revenue between NG and WND. Since the headroom of the generators is shared between regulation and spinning reserve in sharing capacity models, more online reserve requirements are cleared from the cheaper units compared to non-sharing models. The average of the reserve offers for NG are cheaper than WND, approximately half. Therefore, NG earns more ancillary services revenue in S-NC and S-C versus NS-NC and NS-C (Fig. 5 b). It is noteworthy that zero MCPs do not allow this property to show its effect more than this. Due to the large generation capacity, there are remarkable zero MCPs in off-peak intervals in zones one, two, four, and five. Otherwise, the ancillary services revenue gap between sharing and non-sharing models could be extended. Explained impact of capacity sharing helps NG earn more total revenue in S-NC and S-C compared to NS-NC and NS-C (Fig. 5 c).

Different models also influence the revenue distribution between generators in the same category. Normal distributions of energy and ancillary services revenue for NG units are plotted in Fig. 6 to study this impact. A broader bell curve represents the total amount distributed between more units compared to a narrow curve. Cascading or non-cascading requirements do not considerably impact online reserves' revenue distribution (Fig. 6 b). Still, sharing capacity makes the revenue concentrate on the smaller number of NG units because the headroom of the cheaper units can be utilized for regulation and spinning reserves simultaneously, and cheaper units can fulfill both corresponding requirements. However, in non-sharing cases, when the headroom of a unit is used for one of the reserves, the solver must clear the other reserve from another unit. Therefore, a greater number of NG units participate in satisfying the requirements, and online reserves' revenue is distributed between more numbers of units.

On the other hand, sharing capacity does not considerably affect non-spinning revenue (Fig. 6 c). However, cascading requirements make them focus on a smaller number of units. The reason is that a noticeable part of non-spinning requirements is supplied by regulation and spinning products in cascading cases, and only the rest of it is fulfilled by non-spinning. Therefore, fewer units are needed to clear the non-spinning reserve. According to Fig. 6 c, models do not have a noticeable effect on energy revenue distribution.

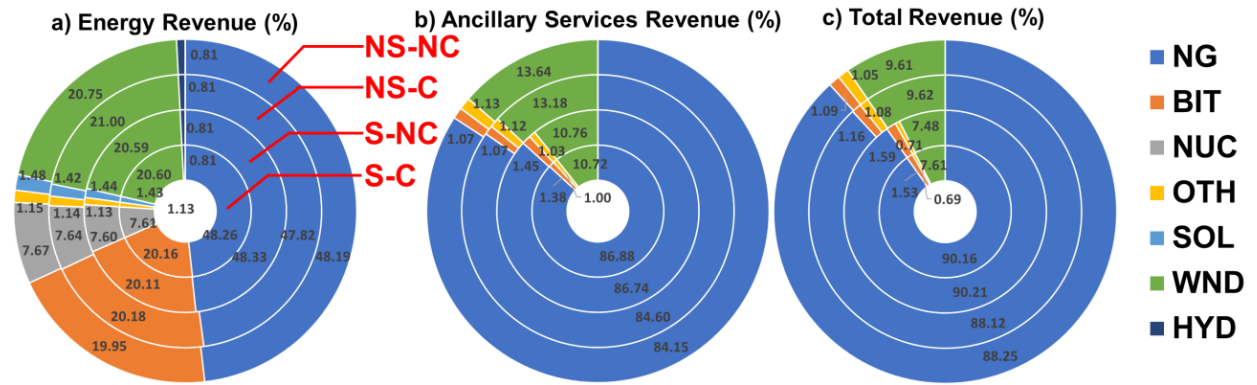

Fig. 5. Share of each type of unit in energy, ancillary services, and total revenue

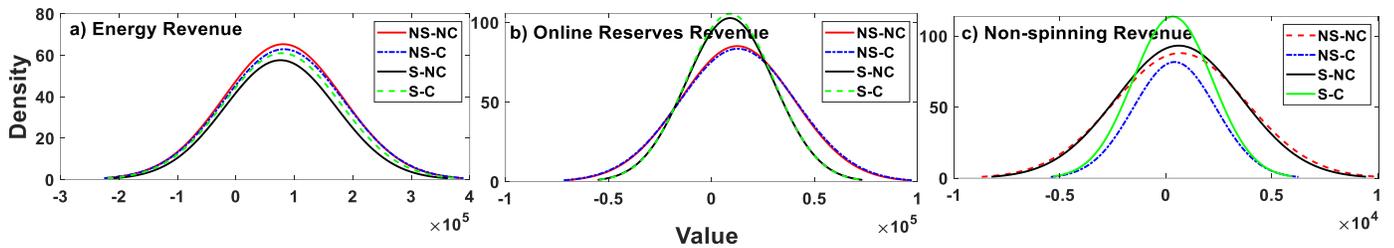

Fig. 6. Normal distribution of energy and ancillary services revenue for natural gas units

## IV. CONCLUSION AND FUTURE WORK

This paper analyzes the influence of the different combinations of the sharing generators' capacity and cascading reserve requirements on market pricing and settlement, considering non-zero ancillary product offers and zonal reserve requirements. The day-ahead electricity market is solved in four models, and MCPs, revenue, costs, and profit are studied in general, individual zones, and diverse unit categories based on fuel type. Remarkable conclusions are as below:

- Zonal MCPs depend on the zonal generation capacity (consequently the proportion of reserve requirements to zonal generation capacity) and reserve offers of market participants at that zone. The higher zonal capacity and the lower offers lead to cheaper MCPs.
- Renewable units with zero fuel costs (SOL, HYD, and WND) have a significant opportunity cost to provide reserve and the most expensive reserve offers. Although they have a very short response time, they are suitable for producing energy rather than clearing reserves. NUC and BIT have the cheapest fuel costs after renewable and relatively expensive reserve offers. Moreover, a very low ramp rate does not allow them to provide a considerable amount of ancillary services, and they also are preferred to provide energy much more than ancillary services. NG supplies most of the reserve requirements due to a high ramp rate, penetration in total capacity, and relatively cheaper offers.
- Sharing capacity shifts the ancillary services' revenue from expensive units (such as renewable) to cheaper units (such as NG) by extending available cheap capacity. Furthermore, sharing capacity constraints make online reserves' revenue concentrate on fewer units compared to non-sharing. On the other hand, cascading reserve requirements do not have a noticeable impact on online reserves' revenue distribution. Yet, it concentrates non-spinning revenue on fewer generators.
- Both cascading reserve requirements and sharing capacity can lower the total system-wide cost by reducing fuel and startup costs. However, sharing capacity is more effective than cascading requirements. Thus, regarding the total cost, NS-NC, NS-C, S-NC, and S-C are ranked first to four.
- Due to less available capacity, non-sharing models lead to more expensive LMPs and MCPs compared to sharing models. As a result, units make more energy, reserve, and total revenue and profit in non-sharing models. Cascading models clear reserves from slightly expensive resources and commit fewer units to make the startup and, consequently, total costs lower. Therefore, they lead to slightly more expensive LMPs and online reserves MCPs, and larger energy, online reserves, and total revenue and profit. Consequently, in these fields, NS-C, NS-NS, S-C, and S-NC are ranked first to four, respectively.
- Regulation and spinning reserves in cascading requirements fulfill a considerable amount of non-spinning requirements. Therefore, marginal units have cheaper offers and non-spinning MCPs are lower. Unlike energy and online reserves, cascading requirements decrease non-spinning revenue and profit.

Different modeling methods of the operating reserve could impact market participants' behavior, including bidding strategies and their response to market signals. As future work, the influence of capacity sharing and reserve substitution on participants' behavior can be investigated. Additionally, capacity sharing reduces generation costs by effectively increasing available capacity and decreasing the reliance on more expensive units. However, this approach might lead to reliability concerns if multiple reserve products need to be deployed simultaneously. The reliability outcomes of various models can be evaluated and monetized accordingly.

This paper focuses on the ancillary service market outcomes under non-scarcity conditions. In the future work, we will examine scarcity pricing of operating reserves under capacity shortage conditions with different reserve formulations and operating reserve demand curve designs.